\documentclass[journal]{IEEEtran}

\usepackage{xcolor,soul,framed} 
\colorlet{shadecolor}{yellow}

\usepackage[pdftex]{graphicx}
\graphicspath{{../pdf/}{../jpeg/}}
\DeclareGraphicsExtensions{.pdf,.jpeg,.png}

\usepackage[]{amsmath}
\usepackage[]{amssymb}
\usepackage{array}
\usepackage{makecell}
\usepackage{pifont}
\usepackage{eqparbox}
\usepackage{url}
\usepackage{etoolbox}
\usepackage{ifthen}
\usepackage{nomencl}
\usepackage[ruled,vlined,linesnumbered]{algorithm2e}

\usepackage{booktabs}    
\usepackage{multirow}    
\usepackage{titlesec}
\usepackage[subtle]{savetrees}
\usepackage{threeparttable, colortbl}
\usepackage{placeins}

\titlespacing*{\section}
{0pt}{6pt}{3pt}

\titlespacing*{\subsection}
{0pt}{4pt}{2pt}

\titlespacing*{\subsubsection}
{0pt}{4pt}{2pt}

\usepackage[colorinlistoftodos]{todonotes}

\newcommand{\MIQCP}{MIQCP}
\newcommand{\MINLP}{MINLP}
\newcommand{\MIBLP}{MIBLP}
\newcommand{\GurobiOff}{\texttt{\sbnb{}-Base}}
\newcommand{\GurobiSBT}{\texttt{\sbnb{}-SBT}}
\newcommand{\GurobiSBTP}{\texttt{\sbnb{}-SBT$^{+}$}}
\newcommand{\PNLP}{\texttt{Penalty-NLP}}
\newcommand{\MILP}{MILP}

\newcommand{\fullMINLP}{Mixed-Integer Nonlinear Program}

\newcommand{\hpmiblp}{$\hat{\textbf{\emph{P}}}_{\text{miblp}}$}
\newcommand{\pmiblp}{$\textbf{\emph{P}}_{\text{miblp}}$}

\newcommand{\sbnb}{sBnB}

\makenomenclature

\usepackage{eso-pic}
\AddToShipoutPictureFG{%
  \AtPageLowerLeft{%
    \put(\LenToUnit{0.5\paperwidth},\LenToUnit{0.35in}){%
      \makebox[0pt][c]{\footnotesize\textit{PREPRINT Version: Submitted to IEEE Transactions on Transportation Electrification}}}}}

\usepackage[colorlinks=true,allcolors=blue]{hyperref}

\begin{document}

\title{Grid-ECO: Grid Aware Electric Vehicle Charging \\ Stations Placement Optimizer}
\author{Bikram Panthee,~\IEEEmembership{Member,~IEEE}, 
        Haoming Yang, 
        Corey D. Harper,
        Amritanshu Pandey~\IEEEmembership{Senior Member,~IEEE}

\thanks{
\noindent B. Panthee and A. Pandey are with the Department of Electrical and Biomedical Engineering, University of Vermont, Burlington, VT 05405, USA
(\{bpanthee, apandey1\}@uvm.edu).\\
H. Yang and C. Harper are with the Department of Civil and Environmental Engineering, Carnegie Mellon University, Pittsburgh, PA 15213, USA
(\{haomingy, cdharper\}@andrew.cmu.edu).
}}

\maketitle

\begin{abstract}
We develop Grid-ECO, a method for optimally allocating electric vehicle charging stations (EVCS) within a distribution feeder while accounting for EV charging demand at census-level granularity.
The underlying problem is a mixed-integer bilinear program (\textbf{MIBLP}) and requires satisfying nonlinear, nonconvex, three-phase unbalanced AC network constraints while including integer siting and sizing decision variables.
Existing works cannot guarantee AC feasibility or optimality without either i) relaxing the integer decision variable space or ii) convexifying AC constraints.
Grid-ECO solves the exact MIBLP to near-zero optimality gap while prioritizing candidate charging locations using grid voltage and current sensitivity metrics. 
To solve the MIBLP exactly, we leverage the global optimization algorithm: spatial branch-and-bound (\sbnb{}). 
To scale the approach to large-scale feeders, we develop a presolving routine that combines i) a penalty-based NLP heuristic for warm-start with b) a sequential bound tightening (SBT) algorithm to derive tight bounds on both bilinear and lifted McCormick variables.
Case studies using realistic Seattle city data demonstrate that Grid-ECO substantially outperforms an off-the-shelf commercial \sbnb{} solver. In three of the four test cases, the commercial solver failed to identify a feasible solution or certify optimality within the 12-hour time limit, whereas Grid-ECO solved all instances to a reported optimality gap of $0.00\%$, with a maximum solver time of 753s. The results further show that tightening both bilinear and lifted McCormick variables reduces \sbnb{} node exploration by up to 98\% and solution time by up to 69\%, while preserving AC-feasible solutions across all test cases.
\end{abstract}


\IEEEpeerreviewmaketitle




\section{Introduction}\noindent The adoption of electric vehicles (EVs) is rapidly increasing, with EV sales in the United States reaching approximately 1.6 million in 2024 and projected to approach 9.6 million by 2030 \cite{IEA2025GlobalEV}.
Supporting this growth would require a substantial expansion of public EV charging infrastructure, which may place significant stress on the existing electric distribution grid \cite{AYABURI2026105333}.
In particular, ad-hoc placement of electric vehicle charging stations (EVCSs) can lead to transformer overloading, excessive voltage drops, and overall grid instability.
Consequently, identifying \textit{where} charging activities occur, along with the corresponding electricity demand \cite{Muratori2020}, becomes a critical design consideration.
This observation motivates the following research problem: \textit{How to identify optimal EVCS locations and charger counts from a set of candidate sites by maximizing charger allocation given a limited budget, grid physics, and census-level charger demand?}
To answer this, we first estimate EV charging demand using a transportation modeling framework based on real-world data such as EV adoption, parking availability, and demographic characteristics. We then use this demand as input to a distribution feeder modeling framework.
After that,  we solve an optimization problem subject to grid physics and budget constraints to satisfy the estimated charging demand and identify the optimal EVCS locations and associated charger counts.

Accurate demand estimation is a prerequisite for EV charging infrastructure planning; however, charging demand is often estimated at a spatial scale that is not directly compatible with feeder-level, grid-aware optimization~\cite{yang2022recent}. Conversely, studies that estimate demand at sufficiently granular spatial resolutions typically do not consider AC power network constraints (e.g.,~\cite{jordan2022electric, liu2024bayesian}).
For example,~\cite{moon2018forecasting} uses discrete-choice experiments and temporal load-shifting models to estimate charging demand at the national level, without accounting for the operational constraints of local distribution networks. Similarly,~\cite{wang2023short} combines high-resolution vehicle trajectory data with deep learning techniques to perform short-term charging demand forecasting, but does not address long-term infrastructure siting and sizing.
More recently,~\cite{yi2023agent} employs an agent-based framework to estimate charging demand at the Traffic Analysis Zone (TAZ) level and optimize service coverage. However, the approach does not incorporate AC network constraints, limiting its applicability to grid-aware EVCS planning.

Granular spatial demand models can estimate EV charger demand at the distribution-feeder level; however, utility planners must ensure that EVCS allocation and charger deployment satisfy grid physics and maintain reliable grid operation. Consequently, EVCS planning requires solving an optimization problem that allocates charging stations and charger capacity while satisfying AC network constraints and meeting charging demand.
The EVCS planning problem is inherently discrete because it involves binary decisions on whether to install a charging station at a candidate location and integer decisions on the number of chargers to deploy. At the same time, AC network constraints introduce nonlinear and nonconvex relationships, resulting in a \fullMINLP{} (\MINLP{}) formulation. The combination of discrete investment decisions and nonlinear power-flow equations makes the problem computationally challenging to solve~\cite{agarwal2022continuous}.

Several studies have incorporated AC power constraints into EVCS planning. A two-stage framework was proposed in~\cite{liu2012optimal} that first screens candidate locations using environmental factors and service-radius requirements and then determines station sizes through a nonlinear AC-constrained optimization model. However, the sequential structure may yield suboptimal solutions because it does not jointly optimize siting and sizing while simultaneously accounting for grid constraints, charging demand, budget limits, and anti-clustering requirements. Moreover, the nonconvex problem is solved using a primal--dual interior-point method, which does not guarantee global optimality.
In~\cite{franco2015mixed}, a three-phase unbalanced AC network model was linearized to formulate an EV charging planning problem. While the resulting model is computationally tractable, the linearization reduces the accuracy of voltage and power-flow representations. Similarly,~\cite{xie2018planning} formulated EVCS planning as a mixed-integer linear program for highway networks, but the model ignores distribution-grid power-flow constraints and assumes that charging demand is supplied exclusively by renewable generators.
In~\cite{cui2019electric}, convex relaxations of AC network constraints were used to minimize the combined cost of charging infrastructure and protection-device upgrades. Likewise,~\cite{zhao2023grid} formulated a mixed-integer second-order cone program that couples transportation and power networks through relaxed AC constraints. However, the study focuses on grid expansion for fast-charging stations along bus routes rather than Level-2 public charging infrastructure in urban distribution systems.
In~\cite{zhu2020grid}, a voltage-impact index was introduced to rank candidate charging-station locations. Although the approach captures voltage sensitivity, it does not explicitly optimize EVCS placement under network operating constraints and therefore may identify locations that are suboptimal or infeasible. In addition, the method considers only voltage impacts and neglects line and transformer loading constraints, which are also critical to distribution-grid reliability.

Despite these contributions, existing solution approaches have gaps. 
They either (i) exclude census-level data, which is required to realistically estimate charger demand and determine locations within the distribution feeder level; (ii) approximate grid physics with linear surrogates; or (iii) relax integer decision variables by their continuous approximations.
\textbf{Grid-ECO} addresses these research gaps by innovating along the following research contributions:
\begin{itemize}
    \item \textbf{Solving \MIBLP{}s to near-zero optimality gap.} 
    Grid-ECO \MIBLP{} formulation includes both unbalanced three-phase AC network constraints and integer decision variables.
    To solve this problem exactly to near-zero optimality gap, we introduce two novel pre-solve mechanisms: i) penalty-based NLP warm-start algorithm, and ii) sequential bound tightening on \emph{both} bilinear and lifted McCormick variables (SBT$^{+}$). 
    With these routines, for downtown and residential Seattle synthetic feeders, Grid-ECO certifies a $0.00\%$ optimality gap with a maximum solver time of $753$\,s, whereas off-the-shelf Gurobi fails to find a feasible solution within $12$~hours.
    
    \item \textbf{Candidate location prioritization.}
    Grid-ECO prioritizes candidate locations using a grid-sensitivity metric that extends the bus voltage sensitivity approach in \cite{zhu2020grid} by incorporating transformer current flow sensitivities. This prioritization also helps to break symmetry while solving the optimization problem.
    
    \item \textbf{Transportation and distribution model integration.} We integrate census block–level EV charging demand derived from a transportation modeling framework as an input into a grid-aware optimization model with exact nonlinear, nonconvex AC distribution network constraints to ensure grid-feasible charger deployment.
    
\end{itemize}

We validate the proposed framework on \textit{two locations (each with two feeders) within the study region, Seattle, WA}, using census block–level charging demand, which is derived from real-world demographic and transportation data and mapped to the distribution feeder. The results show that the Grid-ECO solves  \MIBLP{} to a near-zero optimality gap (see Section~\ref{casestudy}).

\section{Grid-ECO Preliminaries}\label{prelim}
\subsection{Informal Grid-ECO Problem}
\label{sec:grid_eco_framework}
\begin{itemize}

\item[---] \textbf{\textit{Given:}} \textit{Transportation  model and distribution grid model}

\item[---] \textbf{\textit{Output:}} \textit{Optimal EVCS locations and optimal number of chargers at each selected location}

\item[---] \textbf{Key algorithms}: \textit{EV charging demand estimation, candidate location selection, candidate location prioritization, \MIBLP{} optimization routine}
\end{itemize}

The \textit{transportation model} input includes the number of EVs and vehicle registrations, household income, race distribution, and private residential off-street parking availability to estimate the demand for public EVCS in the study region.
The \textit{distribution grid model} input includes the network topology, configuration, and load profiles (including existing EVCS) evaluated at the peak-hour operating point of the feeder. Extension to multi-period or stochastic settings is left to future work. The EV charging demand estimation algorithm (see Section~\ref{sec:charging_demand_modeling}) estimates the net charging demand at the census-block level for the selected study region. 
The candidate location selection and prioritization algorithms (see Section~\ref{prioritize}) identify and prioritize candidate locations. 
The optimization routine (see Sections~\ref{formulation} and~\ref{methodology}) formulates the \MINLP{} and solves it using the proposed methodology.

\subsection{Charging Demand Modeling}
\label{sec:charging_demand_modeling}

Optimal allocation of EVCS depends on charging demand.
We determine the city-wide spatial distribution of EV charging demand using a bi-objective mixed-integer linear programming (MILP) framework, building on work in \cite{lezcano2025siting}. The model outputs the optimal number of charging ports in each census block $i \in \mathcal{B}$ (set of census blocks), denoted by $d_i$.
We sum the net charging demand for all census blocks in the feeder as input for the downstream optimization routine.
The charging demand modeling framework maximizes a weighted combination of utilization efficiency and equitable accessibility, formulated as follows:
\begin{subequations}
\begin{align}
\max_{\mathbf{d}} \quad & (1 - \alpha) \sum_{i \in \mathcal{B}} d_i \mu_i + \alpha \sum_{i \in \mathcal{B}} d_i \epsilon_i \label{eq:obj}\\
\text{s.t.} \quad 
& d_i \leq c_i \quad \forall i \in \mathcal{B}, \label{eq:cap}\\
& \sum_{i \in \mathcal{B}} d_i = b, \label{eq:budget}\\
& d_i \in \mathbb{Z}_{\ge 0} \quad \forall i \in \mathcal{B}. \label{eq:integer}
\end{align}
\end{subequations}
In this formulation, the primary decision variable $d_i$ represents the number of charging ports to be allocated to census block $i$. 
The objective function \eqref{eq:obj} is designed to balance two competing goals. 
The first component is the near-term demand score $\mu_i$, which reflects
expected charger utilization intensity. This score aggregates normalized measures of vehicle density, high-income households, and local EV adoption rates, prioritizing areas with strong indicators for immediate charger use \cite{WANG2026101788}. The second component is the equity need score $\epsilon_i$, which represents the socio-economic necessity for public charging access to support long-term, equitable EV adoption. This score is also constructed using normalized socio-economic indicators, including the proportion of minority and low-income households, as well as the number of households lacking access to off-street parking. The detailed formulations, normalization procedures, and data sources used to construct $\mu_i$ and $\epsilon_i$ are adopted from \cite{lezcano2025siting}. The weighting parameter $\alpha \in [0,1]$ serves as a policy lever to balance the trade-off between prioritizing near-term demand (efficiency) and long-term equity needs (accessibility). In this study, $\alpha = 0.85$ is selected as a neutral reference point, at which the aggregated contributions of the demand and equity components are of comparable magnitude. This reflects a balanced preference between efficiency and equity under the normalized scoring scheme.
The model is bound by two key constraints. 
Constraint \eqref{eq:cap} ensures that the number of charging ports $d_i$ in any given block $i$ does not exceed its transportation infrastructure capacity $c_i$, such as available curbside space. 
Constraint \eqref{eq:budget} enforces that the sum of all allocated chargers across all census blocks equals $b$, 
which represents the total number of chargers to be deployed city-wide, as dictated by a predefined number of charging ports normalized per 1000 households.

\begin{figure}[h]
    \centering
    \includegraphics[width=1\linewidth]{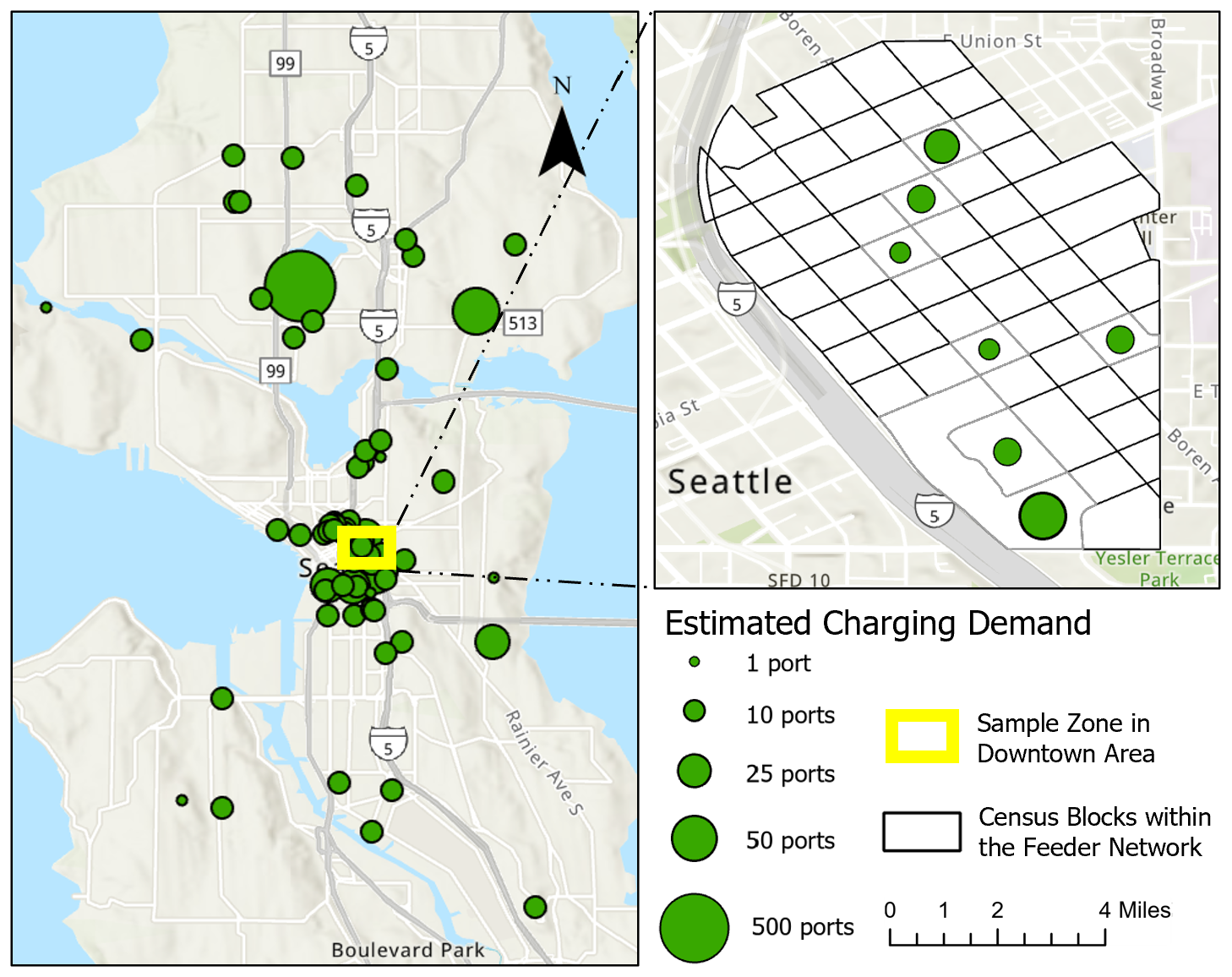}
    \caption{Estimated EV charging port demand in Seattle, WA ($\alpha = 0.85$, 5 ports per 1,000 households). 
    Green circles indicate census-block-level charging demand, with size proportional to the number of ports. 
    The zoomed region shows the selected feeder (D1) in downtown Seattle.}
    \label{fig:seattle_ev_ports}
\end{figure}

The outcome of this optimization is the vector of charger demand for each census block, $\mathbf{d} = [d_1, d_2, \ldots, d_i]$, which represents the spatial distribution of charging demand across census blocks. The total charging demand within the feeder is denoted by $D$, which is computed as the aggregate demand over all census blocks in the selected feeder. Let $\mathcal{F}$ denote the set of census blocks in the selected feeder; then,
\begin{equation}
    D = \sum_{i \in \mathcal{F}} d_i
\end{equation}
We show the output charging demand, for $\alpha=0.85$ for Seattle, WA, in Fig.~\ref{fig:seattle_ev_ports}.
The figure shows the number of new charging ports to be allocated to each census block in the region, given a budget of 5 ports per 1,000 households. 
Additional details of the formulation of the EV siting and sizing module can be found in \cite{lezcano2025siting}.

\subsection{Candidate Location Selection and Prioritization} \label{prioritize}
\begin{figure}[htbp]
    \centering
    \includegraphics[width=1\linewidth]{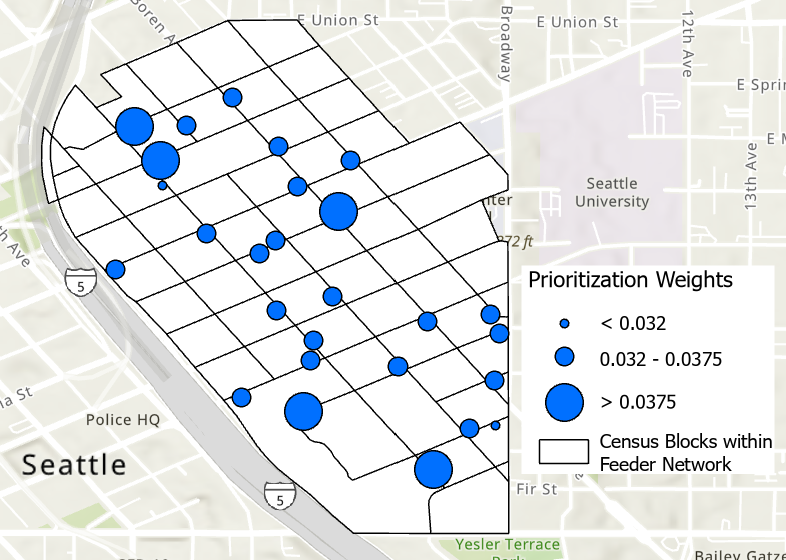}
    \caption{Spatial distribution of candidate locations in feeder (D1) in downtown Seattle, with node sizes proportional to their prioritization weights.}
    \label{fig:gi_index_plot}
\end{figure}

We select candidate locations in the distribution grid based on available transformer capacity. Specifically, candidate locations correspond to pole locations with sufficient remaining transformer capacity. The resulting candidate set is larger than the set of census blocks where charging demand is currently estimated, accounting for areas where charging demand will emerge as EV penetration increases, as illustrated in Fig. \ref{fig:gi_index_plot}. 
\noindent 
We prioritize candidate locations using grid-sensitivity-based weights for two reasons.

\textit{Symmetry breaking.} In combinatorial optimization, locations with similar grid impacts can become nearly interchangeable, causing the \sbnb{} algorithm to explore many equivalent branches. To reduce this effect, we assign distinct weights using~\eqref{softmax} based on how close each location is to grid infeasibility (i.e., a point beyond which grid physics is violated). These weights help break pseudo-symmetry and improve solver performance.

\textit{Grid-impact ranking.} Many distribution feeders operate close to voltage and equipment limits, so additional EVCS demand can lead to constraint violations. We therefore use voltage and transformer-current sensitivities to favor locations with more available grid capacity within the optimization problem in Section \ref{formulation}.

\begin{figure}[h]
    \centering
    \includegraphics[width=1\linewidth]{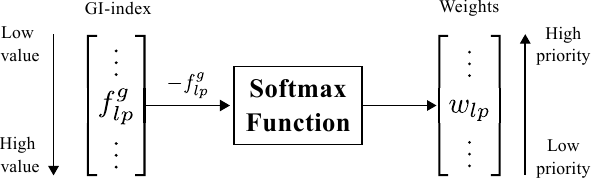}
    \caption{Softmax-based transformation of the Grid Impact Index into priority weights for candidate location selection.}
    \label{fig:prioritze}
\end{figure}

To measure grid sensitivity, we define the Grid Impact Index (GI-index), denoted by $f^g$. A higher GI-index indicates that a location is more sensitive to load injection, while a lower GI-index indicates a better location for EVCS deployment (Fig.~\ref{fig:prioritze}). We compute the GI-index as a weighted combination of the voltage and current impact indices obtained from iterative power-flow simulations. These indices measure the net changes in feeder voltage ($\Delta V$) and current ($\Delta I$) caused by adding one charger at location $(l,p)$. We use the GI-index only to rank candidate locations and construct the prioritization weights $w_{lp}$ in~\eqref{Objective_function}.

\noindent \textbf{Voltage and current impact indices:}
Adapting the work of Zhu \emph{et al.}~\cite{zhu2017voltage}, we define the voltage impact (VI-index), $f^{v}_{lp}$, for a candidate node $l \in \mathcal{L}$ (the set of candidate nodes for EVCS placement) and phase $p \in \Phi$ (with $\Phi = \{A,B,C\}$), as the normalized sum of voltage deviations across all nodes $k \in \mathcal{K}$ (where $\mathcal{K}$ denotes the set of all nodes, and $\mathcal{L} \subseteq \mathcal{K}$) and phases $\phi \in \Phi$, resulting from a unit charger apparent power injection $\delta S_{lp}$ at node $l$ and phase $p$: 
\begin{equation}\label{v_imp_fact}
    f^v_{lp} = \frac{1}{\delta S_{lp}}  \sum_{k \in \mathcal{K}} \hspace{4px} \sum_{\Omega \in \Phi}\Delta V_{k\Omega}^{\delta_{lp}}
\end{equation}
\noindent We compute the net voltage deviation, $\Delta V$, by combining the absolute voltage deviation $\left|\hat{V}_{k\Omega} - V_{k\Omega}\right|$, where $V_{k\Omega}$ and $\hat{V}_{k\Omega}$ denote the voltage magnitude before and after perturbation, with penalty functions $\big(\min(0, \hat{V}_{k\Omega} - V_{k\Omega}^L)$ and $\min(0, V_{k\Omega}^U - \hat{V}_{k\Omega})\big)$ that activate when the perturbed voltage violates the lower and upper voltage limits ($V_{k\Omega}^L$ and $V_{k\Omega}^U$). These violations are scaled by a penalty factor ($\gamma$) to assign higher impact to nodes experiencing limit violations:
\begin{multline}\label{v_dev_fact}
    \Delta V_{k\Omega}^{\delta_{lp}} = \big |(\hat{V}_{k \Omega} - V_{k \Omega}) \big|\\ + \gamma \big| \left ( \min(0, \hat{V}_{k\Omega} - V_{k \Omega}^L) + \min(0, V_{k\Omega}^U - \hat{V}_{k\Omega}) \right ) \big|
\end{multline}

\noindent Similarly, we define the CI-index ($f^c_{lp}$) as the sum of net current deviations $\Delta I_{\tau}^{\delta_{lp}}$ across all transformers $\tau \in \mathcal{T}$, where $\mathcal{T}$ denotes the set of transformers:
\begin{equation}\label{c_imp_fact}
    f^c_{lp} = \frac{1}{\delta S_{lp}} \sum_{\tau \in \mathcal{T}} \Delta I_{\tau}^{\delta_{lp}}
\end{equation}
where,
\begin{equation}
    \Delta I_{\tau}^{\delta_{lp}} = \left( |(\hat{I}_{\tau} - I_{\tau})| + \gamma \big |\min(0, I_\tau^{\text{rat}} - \hat{I}_{\tau}) \big | \right)
\end{equation}
where $I_{\tau}$ and $\hat{I}_{\tau}$ denote the transformer current flow before and after perturbation, respectively. $I_{\tau}^{\text{rat}}$ denotes the current rating of transformer $\tau$.

\noindent We stack the computed VI-index and CI-index into the VI-index vector $f^v$ and the CI-index vector $f^c$, for all locations $(l, p)$, respectively.

\medskip

\noindent \textbf{Grid impact index calculation:}
To ensure unit consistency, we normalize the vectors of the voltage and current indices, $f^v$ and $f^c$, by their respective maximum values: $f^{n,v}_{lp} = f^{v}_{lp}/\max(f^v)$ and $f^{n,c}_{lp} = f^{c}_{lp}/\max(f^c)$. We then compute the GI-index as:
\begin{equation}\label{g_index}
    f^g_{lp} = a_{lp} \cdot f^{n,v}_{lp} + b_{lp} \cdot f^{n,c}_{lp}
\end{equation}
\noindent \noindent We compute the weights $a_{lp}$ and $b_{lp}$ in~\eqref{weight_gi} proportionally so that the dominant impact determines the index to produce location-specific
weights.
\begin{equation}\label{weight_gi}
    a_{lp} = \frac{f^{n,v}_{lp}}{f^{n,v}_{lp}+f^{n,c}_{lp}}, \quad b_{lp} = \frac{f^{n,c}_{lp}}{f^{n,v}_{lp}+f^{n,c}_{lp}}
\end{equation}
\noindent \textbf{Weight Assignment:}
Finally, to favor locations with low grid impact, we assign prioritization weights $w_{lp}$, for all locations $(l,p)$, using the negative softmax function on the GI-index. This approach, illustrated in Fig. \ref{fig:gi_index_plot}, assigns higher weights to lower GI-index values:
\begin{equation}\label{softmax}
w_{lp}
=
\operatorname{softmax}\!\left(-f^{g}_{lp}\right)
=
\frac{\exp\!\left(-f^{g}_{lp}\right)}
{\displaystyle \sum_{j \in \mathcal{L}} \sum_{\phi \in \Phi}
\exp\!\left(-f^{g}_{j\phi}\right)}
\end{equation}


\section{Grid-ECO Optimization Model Formulation}\label{formulation} 
\noindent Given feeder EV chargers demand, and candidate locations with priority weights, we formulate a constrained \MIBLP{} optimization problem~\eqref{MIBLP}  to find the optimal locations and the number of chargers.
\subsection{Objective Function}
In this formulation, the utility planner operates under a fixed budget and limited installation resources, and therefore seeks to maximize the overall deployment value within the available investment constraints. Grid-ECO maximizes this deployment value by improving user access to EVCS through the installation of the maximum number of chargers across prioritized candidate locations while satisfying \textit{at least} the charging demand $D$.
Mathematically, the objective function is:
\begin{equation}\label{Objective_function}
    f(\mathbf{x}, \mathbf{z})  = \sum_{l \in \mathcal{L}} \hspace{4px} \sum_{p \in \Phi} w_{lp}\cdot x_{lp} \cdot z_{lp}
\end{equation}
\noindent
The parameter $w_{lp}$ denotes the priority weight associated with installing chargers at candidate location $l \in \mathcal{L}$, where $\mathcal{L}$ is the set of candidate EVCS nodes, and phase $p \in \Phi$, with $\Phi = \{A,B,C\}$. The binary decision variable $x_{lp} \in \{0,1\}$ equals $1$ if a charging station is installed at location $(l,p)$. The nonnegative integer variable $z_{lp}$ represents the number of chargers installed at location $(l,p)$.

\subsection{Constraints}
\subsubsection*{\textbf{Grid Physics Constraint}}
We enforce grid physics by placing charging stations to satisfy the Kirchhoff current law (KCL)-based AC network constraint at every node. 
To model AC network constraints, we use a three-phase equivalent circuit \cite{pandey2018robust} and express power flow with the current-voltage ($IV$) formulation. 
We use the $IV$ formulation over the $PQV$ formulation because it allows for \textit{exact} bilinear formulation of the problem.
For readability, we exclude three-phase transformers, shunts, switches, regulators, and fuses from the notation, but include them in the 
implementation.
We model the complex power injection $(\widetilde{S})$ for loads and chargers as a product of complex voltage $(\widetilde{V})$ and complex current $(\widetilde{I})$:
\begin{equation} \label{S_vi_eqn}
    \widetilde{S} = P + jQ = \widetilde{V} \widetilde{I}^*
\end{equation}
The three-phase AC network constraints in rectangular form are given by  \eqref{real_kcl}--\eqref{Ii_line}.

\noindent For all $p \in \Phi$ and $k \in \mathcal{K}$:
\begin{subequations}\label{grid_constraints}
\begin{alignat}{1}
         & I_{kp}^{r,load} + I_{kp}^{r,line} + I_{kp}^{r,ch} = 0 \label{real_kcl}\\
         & I_{kp}^{i,load} + I_{kp}^{i,line} + I_{kp}^{i,ch} = 0 \label{imag_kcl}\\
         & P_{kp}^{load} - \left ( V_{kp}^r I_{kp}^{r,load} + V_{kp}^i I_{kp}^{i,load} \right ) = 0 \label{P_load} \\
         & Q_{kp}^{load} - \left ( V_{kp}^i I_{kp}^{r,load} - V_{kp}^r I_{kp}^{i,load} \right ) = 0\label{Q_load} \\
         & P_{kp}^{ch} - \left ( V_{kp}^r I_{kp}^{r,ch} + V_{kp}^i I_{kp}^{i,ch}\right ) = 0 \label{P_ch} \\
         & Q_{kp}^{ch} -\left ( V_{kp}^i I_{kp}^{r,ch} - V_{kp}^r I_{kp}^{i,ch}\right ) = 0 \label{Q_ch} \\
         & I_{kp}^{r,line} - \left (\sum_{j \in \mathcal{K}} \sum_{\phi \in \Phi}G^{line}_{kj, p \phi} V_{kj, \phi}^r - B^{line}_{kj, p \phi}V_{kj,\phi}^i \right)= 0\label{Ir_line} \\
         & I_{kp}^{i,line} - \left (\sum_{j \in \mathcal{K}} \sum_{\phi \in \Phi}G^{line}_{kj, p\phi } V_{kj, \phi}^i + B^{line}_{kj, p \phi}V_{kj,\phi}^r \right)= 0\label{Ii_line}
\end{alignat}
\end{subequations}

\noindent Here, $I_{kp}^{r,load}$ and $I_{kp}^{i,load}$ denote the real and imaginary load currents, while $I_{kp}^{r,ch}$ and $I_{kp}^{i,ch}$ denote the real and imaginary charging currents leaving the node $k \in \mathcal{K}$ (where $\mathcal{K}$ denotes the set of all nodes) and phase $p$, respectively. Similarly, $I_{kp}^{r,line}$ and $I_{kp}^{i,line}$ represent the net real and imaginary line currents flowing out of node $k$ and phase $p$.
$V_{kp}^r$ and $V_{kp}^i$ are the real and imaginary voltage variables at node $k$ and phase $p$.
The parameters $P_{kp}^{load}$ and $Q_{kp}^{load}$ represent the peak active and reactive power demand \textit{not including} EVCS at node $k$ and phase $p$.
The variables $P_{kp}^{ch}$ and $Q_{kp}^{ch}$ denote the active and reactive power demand by EVCS at node $k$ and phase $p$.
$G^{line}_{kj, p \phi}$ and $B^{line}_{kj, p \phi}$ denote the conductance and susceptance parameters from the multi-phase admittance matrix, respectively.
Equations \eqref{real_kcl} and \eqref{imag_kcl} enforce KCL at each node $k$ and phase $p$. 
The charger currents $I_{kp}^{r,ch}$ and $I_{kp}^{i,ch}$ activate only if $k \in \mathcal{L}$ (a candidate EVCS location); otherwise, they are zero. 
Equations \eqref{P_load}, \eqref{Q_load}, \eqref{P_ch}, and \eqref{Q_ch} represent the bilinear active and reactive power through loads and chargers, stemming directly from the definition in \eqref{S_vi_eqn}. 
Equations \eqref{Ir_line} and \eqref{Ii_line} model the linear line currents.

\subsubsection*{\textbf{Grid Limit Constraint}}
We enforce grid operational limits to maintain system reliability. Specifically, we constrain node voltages within specified bounds and ensure that transformer and line currents remain below their thermal ratings. We model other equipment-specific limits similarly.

\noindent \underline{Voltage limit}:
\begin{equation}\label{voltage_limit}
   \left (V_{kp}^L\right)^2 \le \left (V_{kp}^r\right )^2 + \left (V_{kp}^i \right)^2 \le \left (V_{kp}^U\right )^2, \quad \forall p \in \Phi, \forall k \in \mathcal{K}
\end{equation}

\noindent \underline{Transformer current limit}:
\begin{equation}\label{current_limit}
     \left( I_{\tau}^{r} \right)^2 + \left( I_{\tau}^{i} \right)^2 \le \left (I_{\tau}^{rat}\right )^2, \quad \forall \tau \in \mathcal{T} 
\end{equation}
Here, $V_{kp}^{L}$ and $V_{kp}^{U}$ represent the lower and upper bounds on the voltage magnitude at node $k$ and phase $p$, respectively. $I_{\tau}^{r}$ and $I_{\tau}^{i}$ denote the real and imaginary components of the current flowing through transformer $\tau$, while $I_{\tau}^{\text{rat}}$ denotes the current rating of transformer $\tau$.

\subsubsection*{\textbf{Charger Demand Constraint}}
We want to ensure that at least $D$ number of chargers are installed in the selected feeder. 
\begin{equation}\label{power_demand_constraint}
        D - \sum_{l \in \mathcal{L}} \hspace{4px} \sum_{p \in \Phi}  z_{lp} \le 0
\end{equation}
Furthermore, parking availability is limited; therefore, constraint~\eqref{xz_constraint} ensures that charger deployment respects physical limits, as defined by the minimum and maximum allowable chargers, $z_{lp}^L$ and $z_{lp}^U$.
\begin{subequations}\label{xz_constraint}
    \begin{alignat}{4}
        & x_{lp} z_{lp}^L - z_{lp} \le 0, \quad \forall p \in \Phi,\forall l \in \mathcal{L}\\
        & z_{lp} - x_{lp} z_{lp}^U\le 0, \quad \forall p \in \Phi,\forall l \in \mathcal{L}
    \end{alignat}
\end{subequations}
Since $z_{lp}$ is a nonnegative integer, constraint \eqref{xz_constraint} also ensures that if the location is not selected $(x_{lp} = 0)$ for placement of a charging station, then no chargers are installed at that location.

The active power consumed by an EVCS equals the total power drawn by all chargers deployed at that location, given by the number of installed chargers multiplied by the charger power rating $P^{ch,\text{rat}}$:
\begin{equation} \label{ch_rating}
    P_{lp}^{ch} - z_{lp} P^{ch,\text{rat}} = 0
\end{equation}

\noindent Additionally, we want to enforce power factor $(\text{pf})$ for all EVCS:
\begin{equation}\label{pf_constraint}
    Q_{lp}^{ch} - P_{lp}^{ch} \cdot \tan{(\cos^{-1}{(\text{pf})})} = 0, \quad \forall p \in \Phi, \forall l \in \mathcal{L}
\end{equation}

\subsubsection*{\textbf{Budget Constraint}}
We ensure that the total cost, including land acquisition ($C_{lp}^{CS}$) and charger installation ($C^{ch}$), remains within the available budget ($C$).
\begin{equation}\label{budget_constraint}
    \sum_{l \in \mathcal{L}} \hspace{4px} \sum_{p \in \Phi} \left (C_{lp}^{CS}x_{lp} +  C^{ch} z_{lp} \right ) \le C
\end{equation}

\subsubsection*{\textbf{Anti-Clustering Constraint}}
\begin{figure}[h]
    \centering
    \includegraphics[width=1\linewidth]{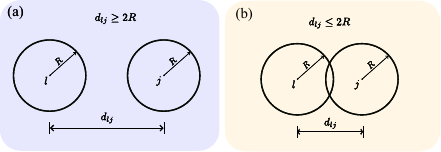}
    \caption{Illustration of the service radius for two charging stations. In (a), the service radii do not intersect, and the constraint is inactive. In (b), the service radii intersect, activating the anti-clustering constraint.}
    \label{fig:service_radius}
\end{figure}

Since the objective (Section~\ref{Objective_function}) maximizes user access by maximizing the number of stations and chargers, charging stations may cluster in the same region. We prevent this by enforcing spatial separation among candidate locations using a predefined service radius $R$.
As shown in Fig.~\ref{fig:service_radius}, if the service areas of two candidate locations overlap, i.e., if the distance between locations $d_{lj}$ is less than twice the service radius $2R$, then we allow at most one charging station at those locations. We enforce this condition using the following constraint:
\begin{equation}\label{radius_constriant}
    x_{l} + x_{j} \leq 1, \quad \text{if } l \neq j \text{ and } d_{lj} \leq 2R, \quad \forall l, j \in \mathcal{L}.
\end{equation}
We compute the service radius $R$ following \cite{liu2012optimal}, where it is defined as the maximum distance an EV user is willing to travel to reach a charging station. To evaluate the distance between two candidate locations, we use the Haversine formula \cite{mohammed2022internet}, which accounts for the Earth's curvature and computes distances based on latitude and longitude coordinates. The Haversine distance is given by:
\begin{align}
    &a = \sin^2 \left(\frac{\Delta \text{lat}}{2} \right) + \cos(\text{lat}_l)\cos(\text{lat}_j)\sin^2 \left(\frac{\Delta \text{long}}{2} \right), \\
    &c = 2 \cdot \text{atan2} \left( \sqrt{a}, \sqrt{1 - a} \right), \\
    &d_{lj} = r \cdot c,
\end{align}
where $\text{lat}_l$ and $\text{lat}_j$ denote the latitudes of locations $l$ and $j$, respectively, and $\text{long}_l$ and $\text{long}_j$ denote their longitudes. Here, $\Delta \text{lat} = \text{lat}_j - \text{lat}_l$ and $\Delta \text{long} = \text{long}_j - \text{long}_l$. The parameter $r$ represents the mean radius of the Earth, approximately $6{,}371$~km.
By combining all constraints and the objective function, we formulate a Mixed-Integer Bilinear Program (MIBLP):
\begin{subequations}\label{MIBLP}
    \begin{align}
        \textbf{\emph{P}}_{\text{miblp}}:\max\quad & f(\mathbf{x}, \mathbf{z}) \label{MIBLP_a} \\
        \text{s.t.}\quad & g(\mathbf{y}, \mathbf{s}) = 0 \label{MIBLP_b} \\
                         & h(\mathbf{x}, \mathbf{y}, \mathbf{z},\mathbf{s}) \le 0 \label{MIBLP_c} \\
                         & \mathbf{y}^L \le \mathbf{y} \le \mathbf{y}^U \label{MIBLP_d}\\
                         & s_k = y_iy_j \quad \forall k\in BL \label{MIBLP_e}\\
                         & \mathbf{x} \in \{0,1\}^{n_x} \label{MIBLP_f} \\
                         & \mathbf{z} \in \mathbb{Z}_{\ge 0}^{n_x} \label{MIBLP_g}
    \end{align}
\end{subequations}
\noindent In \eqref{MIBLP}, the objective function \eqref{MIBLP_a} is defined by \eqref{Objective_function}.
Constraint \eqref{MIBLP_b} represents the grid physics constraints in \eqref{real_kcl}--\eqref{Ii_line}, where $\mathbf{y}$ denotes the vector of all AC network state variables.
$s_k$ denotes the $k^{\text{th}}$ bilinear term, where each $s_k = y_i y_j$ for $(i, j) \in BL$, and the vector $\mathbf{s}$ contains all such bilinear terms, i.e., $\mathbf{s} = \{s_k \mid k \in BL\}$. The equality constraint \eqref{MIBLP_b} includes the linear and bilinear AC network constraints.
Constraint \eqref{MIBLP_c} captures the inequality constraints, including grid operating limits in \eqref{voltage_limit}--\eqref{current_limit}, charger demand constraints in \eqref{power_demand_constraint}--\eqref{pf_constraint}, the budget constraint in \eqref{budget_constraint}, and the service radius constraint in \eqref{radius_constriant}.
$\mathbf{x}$ is a vector of binaries, which describes whether a charging station should exist at a location or not, and  $ \mathbf{z}$ models the number of chargers deployed at those locations.
Constraint \eqref{MIBLP_d} bounds the state variables.
The AC network equations are bilinear equality constraints and hence non-convex \eqref{MIBLP}. 
The decision variables governing EVCS placement and charger counts in \eqref{MIBLP_e} and \eqref{MIBLP_f} are integers and also increase the complexity of the formulation, resulting in a non-convex \MIBLP{}.

Current works, including \cite{franco2015mixed, xie2018planning, cui2019electric}, do not solve exact \pmiblp{} and either (i) relax the nonlinear AC network constraints into its convex surrogates while retaining the discrete variables and solving the resulting \MILP{}, or (ii) relax the discrete variables to obtain an NLP that can be solved using local solvers.
However, we cannot certify the quality of these solutions, as they may be infeasible in the AC network paradigm in case (i) and may be sub-optimal in case (ii).

Recent advancements in global solvers such as Gurobi and BARON can solve \MIBLP{}s, to global or near-global optimality, as measured by the optimality gap. For \MIBLP{}s, these solvers use the spatial branch-and-bound (\sbnb{}) algorithm~\cite{smith1996global} to handle nonconvexities.
In our previous work~\cite{panthee2025solving}, we solve the power grid bilinear problem (BLP) to a near-zero optimality gap using global solvers.
Building on that work, this paper investigates the effectiveness of the \sbnb{} algorithm for solving bilinear optimization problems with discrete decision variables, i.e., \MIBLP{}s, to near-zero optimality gap.
We find the methods in \cite{panthee2025solving} alone are unable to solve \MIBLP{} problems to global optimality in a practical amount of time.

\section{Spatial Branch and Bound (sBnB) Algorithm}
The \sbnb{} algorithm solves the \pmiblp{} problem by recursively partitioning the solution space into smaller sub-spaces and solving a sequence of relaxed subproblems~\cite{smith1996global}.
During this process, the algorithm prunes subspaces that cannot improve the incumbent solution. 
For each node in the search tree, the solver applies convex relaxations and bound-tightening procedures, solves the relaxed problem, and attempts to recover an integer- and nonlinear-feasible solution through a standard branch-and-bound procedure.
The algorithm terminates when the incumbent objective value and the best bound coincide, or when their relative difference falls below an optimality tolerance.
For a maximization problem, the incumbent provides a lower bound on the global optimum, whereas the best bound provides an upper bound.

\begin{figure}[h]
    \centering
    \includegraphics[width=1\linewidth]{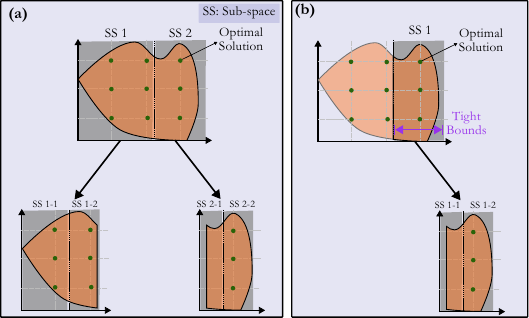}
    \caption{The figure shows the branching process of the \sbnb{} algorithm. Fig. (a) shows the traditional method, which explores all branches. Fig. (b) shows the Grid-ECO method, where the algorithm explores fewer branches.}
    \label{fig:sbb}
\end{figure}

Although modern global solvers can potentially solve large-scale \MIBLP{}s, their practical performance is often not robust. We call this direct application of the off-the-shelf solver, \GurobiOff{}, and Table~\ref{tab:comparison} shows its poor performance across all test cases.
The challenge is on two fronts: i) finding a good incumbent solution and ii) exploring the large search space.
An incumbent is a feasible solution to the optimization problem. Global solvers often struggle to recover incumbents quickly because they must satisfy both integer and nonlinear constraints simultaneously. As shown in Table~\ref{tab:comparison}, \GurobiOff{} failed to find an incumbent within the allotted time for 3 out of the 4 test cases.
Global solvers have in-built heuristics that can improve feasibility detection, but these heuristics often consume substantial computational time and produce incumbents that may be far from the global optimal solution.
The second challenge depends on the size of the search space and directly correlates to the initial or processed bounds on the decision and dependent variables.
When explicit bounds are unavailable, global solvers such as Gurobi derive them through presolve procedures; however, these bounds are often weak. 
Loose bounds weaken the convex outer approximations and force the \sbnb{} algorithm to explore large and uninformative regions of the feasible space, as illustrated in Fig.~\ref{fig:sbb}(a).
Consequently, the computational performance of off-the-shelf solvers degrades substantially for large-scale \MINLP{}s~\cite{das2024branch}.

\section{Grid-ECO Solution Methodology}\label{methodology}
\begin{figure}[h]
    \centering
    \includegraphics[width=1\linewidth]{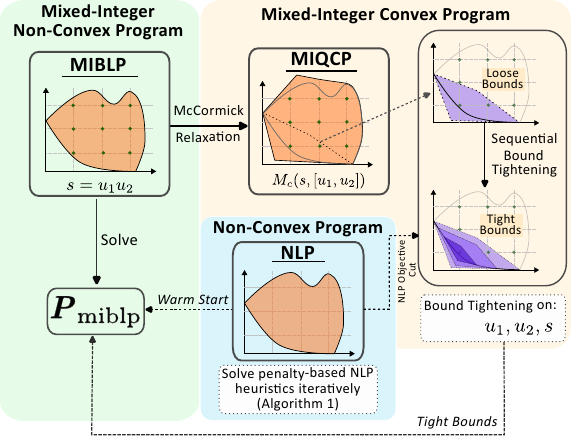}
    \caption{Overview of the Grid-ECO solution methodology for solving \pmiblp{} to near-zero optimality gap.}
    \label{fig:opt_engine}
\end{figure}

We introduce Grid-ECO to improve the performance of the \sbnb{} algorithm for \MIBLP{}, focusing on two presolve strategies (see Fig. \ref{fig:opt_engine}).
First, Grid-ECO builds a pre-solve routine that provides locally optimum feasible warm starts to serve as strong incumbents early in the optimization process, as discussed in Subsection~\ref{nlp_heuristics}.
Second, Grid-ECO uses the sequential bound tightening framework to derive tight bounds for the bilinear variables and the lifted McCormick variables in Subsection~\ref{cvx_relax_bound_tight} to shrink the feasible search space, strengthen the convex relaxations, and improve the convergence of the \sbnb{} algorithm, as illustrated in Fig.~\ref{fig:sbb}(b).

\subsection{Penalty-Based Warm Start} \label{nlp_heuristics}
\begin{algorithm}[!htbp]
\caption{\textbf{Penalty-Based Warm-Start (NLP)}}
\label{alg:penalty_heuristic}
\KwIn{penalty weights ($\mu = \mu_0$, $\eta = \eta_0$, $\mu_0 > \eta_0$), growth factor $\lambda > 1$, max iterations $iter$, tolerances $\epsilon_x, \epsilon_z$.}
\textbf{Relax:} $x_{lp} \in [0,1]$ and $z_{lp} \in [0, z_{lp}^U]$.

\For{$i = 1, 2, \dots, iter$}{
    \textbf{Solve NLP:}
    \begin{subequations}\label{NLP}
    \begin{align}
        \max \hspace{2px}
        & f(\mathbf{x}, \mathbf{z}) -\mu \sum_{l \in \mathcal{L}} \hspace{2px} \sum_{p \in \Phi}t_{lp}^x - \eta \sum_{l \in \mathcal{L}} \hspace{2px} \sum_{p \in \Phi} t_{lp}^z \label{NLP_a} \\[2pt]
        \text{s.t.}\quad 
        & \eqref{MIBLP_b}, \eqref{MIBLP_c}, \eqref{MIBLP_d}, \eqref{MIBLP_e} \label{NLP_b} \\
        & x_{lp}(1 - x_{lp}) \le t_{lp}^x, \quad \forall l \in \mathcal{L}, p \in \Phi \label{NLP_c} \\
        &  \sin^2(\pi z_{lp}) \le t_{lp}^z, \quad \forall l \in \mathcal{L}, p \in \Phi \label{NLP_d} \\
        & t_{lp}^x \ge 0, t_{lp}^z \ge 0, \quad \forall l \in \mathcal{L}, p \in \Phi \label{NLP_e}
    \end{align}
\end{subequations}

    \textbf{Compute Gaps:}

    $\Delta_x = \max |\mathbf{x} - \text{round}(\mathbf{x})|$; \hspace{4px}
    $\Delta_z = \max |\mathbf{z} - \text{round}(\mathbf{z})|$\;
    \If{$\Delta_x \le \epsilon_x$ \textbf{and} $\Delta_z \le \epsilon_z$}{
        Integer-feasible solution found; \textbf{break}
    }

    \textbf{Update Weights:}

    \If{$\Delta_x \le \epsilon_x$ \textbf{and} $\Delta_z > \epsilon_z$}{
        $\mu \gets \mu \cdot \lambda$; 
        $\eta \gets \eta \cdot \lambda^2$ 
    }
    \Else{
        $\mu \gets \mu \cdot \lambda$; \quad
        $\eta \gets \eta \cdot \lambda$
    }
    \textbf{Initialize:} 
    $\mathbf{x}_{i+1} \gets \mathbf{x}_{i}$; \quad $\mathbf{z}_{i+1} \gets \mathbf{z}_{i}$\;
}
\textbf{Post-processing:} Fix $\hat{\mathbf{x}} \gets \text{round}(\mathbf{x})$ and $\hat{\mathbf{z}} \gets \text{round}(\mathbf{z})$;
Assign AC network variables $\hat{\mathbf{y}} \gets \mathbf{y}$
\end{algorithm}

We implement a penalty-based warm-start presolve heuristic to obtain an integer- and nonlinear-feasible solution for the \pmiblp{} problem. 
The resulting solution yields a high-quality incumbent solution, which serves as a tight primal bound, enabling the \sbnb{} solver to prune the search space more aggressively. 
Algorithm~\ref{alg:penalty_heuristic} describes the warm-start presolve routine.
We first relax the binary variables $x_{lp}$ and integer variables $z_{lp}$ to $x_{lp} \in [0,1]$ and $z_{lp} \in [0,z_{lp}^U]$, respectively, and subsequently solve a sequence of continuous nonlinear programs (NLPs) following a homotopy approach.
At each iteration, Algorithm~\ref{alg:penalty_heuristic} solves the penalized NLP in~\eqref{NLP}, where integrality is enforced through two penalty mechanisms tailored to the corresponding variable types. 
For the binary variables $x_{lp}$, we employ the quadratic penalty function $x_{lp}(1-x_{lp})$ in~\eqref{NLP_c}, which evaluates to zero if and only if $x_{lp} \in \{0,1\}$~\cite{ali2025ac}. 
For the nonnegative integer variables $z_{lp}$, we adopt the periodic sinusoidal penalty function $\sin^2(\pi z_{lp})$ in~\eqref{NLP_d}, following the approach in~\cite{shin1990penalty}. This function vanishes at every integer point and therefore naturally penalizes fractional values.

Rather than incorporating these penalties directly into the objective function, we impose them through inequality constraints using nonnegative slack variables, $t$. 
This formulation preserves the smoothness of the objective function and improves numerical conditioning by isolating the oscillatory penalty terms from the evaluation of the objective. 
Since the \pmiblp{} formulation represents a maximization problem, the objective in~\eqref{NLP_a} subtracts the weighted slack penalties $\mu \sum t_{lp}^x + \eta \sum t_{lp}^z$, thereby driving both slack variables toward zero and simultaneously enforcing integrality in $x_{lp}$ and $z_{lp}$. 
The penalty parameters $\mu> \eta>0$ regulate the relative emphasis placed on binary and integer feasibility, respectively. After each NLP solve, we evaluate the integrality violations using $\Delta_x$ and $\Delta_z$, defined as the maximum absolute deviation of the variables from their nearest integer values.

If the NLP solution remains fractional, the algorithm updates the penalty parameters and warm-starts the subsequent NLP using the current solution. When both $x_{lp}$ and $z_{lp}$ violate integrality, the algorithm scales both $\mu$ and $\eta$ by the growth factor $\lambda>1$, thereby uniformly strengthening the penalty enforcement. When the binary variables satisfy integrality, i.e., $\Delta_x \leq \epsilon_x$, but the integer variables remain fractional, i.e., $\Delta_z > \epsilon_z$, the algorithm increases $\eta$ more aggressively using $\lambda^2$ while scaling $\mu$ by $\lambda$. This adaptive update strategy concentrates the penalty increase on the harder-to-round integer variables. In all cases, the algorithm initializes $\mathbf{x}_{i+1} \leftarrow \mathbf{x}_i$ and $\mathbf{z}_{i+1} \leftarrow \mathbf{z}_i$, thereby exploiting the continuity of successive NLP iterates and avoiding repeated cold starts. Upon termination, the algorithm rounds the final solution to the nearest integer point.

\subsection{Bound Tightening}
\label{cvx_relax_bound_tight}

As a second presolve routine, we introduce and apply the Sequential Bound Tightening (SBT) method proposed in~\cite{nagarajan2016tightening}, which iteratively solves the bound contraction algorithm introduced in~\cite{CASTRO2015300} to obtain tight variable bounds.
By tightening variable bounds, SBT shrinks the feasible region of the optimization problem and strengthens the convex relaxations used in the \sbnb{} algorithm, thereby improving convergence and reducing the number of explored nodes.

To implement SBT, we first construct convex outer approximations of all bilinear terms using the McCormick relaxation~\cite{mccormick1976computability}. 
The resulting relaxation~\eqref{sbt_equations} is then solved iteratively.
For a generic bilinear term $s=u_1u_2$, the McCormick envelope operator $M_c(\cdot)$ is defined as
\begin{equation}\label{mccormick}
    M_c(s,[u_1,u_2])=
    \begin{cases}
        u_1^Uu_2 + u_1u_2^U - u_1^Uu_2^U - s \le 0,\\
        u_1^Lu_2 + u_1u_2^L - u_1^Lu_2^L - s \le 0,\\
        s - u_1^Uu_2 - u_1u_2^L + u_1^Uu_2^L \le 0,\\
        s - u_1^Lu_2 - u_1u_2^U + u_1^Lu_2^U \le 0,
    \end{cases}
\end{equation}
where $u_1^L$, $u_1^U$, $u_2^L$, and $u_2^U$ denote the lower and upper bounds of the bilinear variables.

We extend standard SBT by tightening bounds on both the original bilinear variables $(u_1,u_2)$ and the associated lifted McCormick variables $s$. Unlike conventional SBT approaches that focus only on the original bilinear variables, our approach also tightens lifted McCormick variables and propagates the resulting bounds back to the original formulation. To the best of the authors' knowledge, this strategy has not been previously applied to three-phase unbalanced AC EVCS placement and sizing problems.

For example, consider the bilinear term $V_{kp}^r I_{kp}^r$ in~\eqref{P_load}.
To apply SBT, we introduce the lifted variable $s_{kp}^1 = V_{kp}^r I_{kp}^r$ and construct the corresponding McCormick envelope $M_c(s_{kp}^1,[V_{kp}^r,I_{kp}^r])$.
The SBT method then tightens the bounds of $V_{kp}^r$, $I_{kp}^r$, and $s_{kp}^1$ simultaneously. 
Tightening the bounds of the lifted variable directly yields substantially stronger bounds than propagating the bounds of $V_{kp}^r$ and $I_{kp}^r$ through interval arithmetic because interval arithmetic treats repeated variable occurrences independently and therefore cannot exploit the dependency structure embedded in the bilinear relation~\cite{hickey2001interval}.
After obtaining the tightened bounds, we include them in the \pmiblp{} in lieu of the original loose bounds.
Results in Table~\ref{tab:comparison} show that \GurobiSBTP{} (SBT on both bilinear and lifted McCormick variables) consistently explores fewer \sbnb{} branches than \GurobiSBT{} (SBT applied only to bilinear variables).

\begin{algorithm}[!htbp]
\caption{\textbf{SBT on $\boldsymbol{\omega}$ (MIQCP)}}
\label{alg:sbt}
\KwIn{original loose $\boldsymbol{\omega}^L$, $\boldsymbol{\omega}^U$, tolerance $\epsilon$}

\textbf{Solve:} \pmiblp{} using algorithm \ref{alg:penalty_heuristic} to obtain 
$f_{\text{nlp}}(\mathbf{x}^*, \mathbf{z}^*)$

\textbf{Initialize:} $\boldsymbol{\omega}^l \leftarrow \boldsymbol{\omega}^L$,\quad 
$\boldsymbol{\omega}^u \leftarrow \boldsymbol{\omega}^U$

\textbf{Set:} $\boldsymbol{\omega}^l_{\text{prev}} \leftarrow \mathbf{0}$,\quad 
$\boldsymbol{\omega}^u_{\text{prev}} \leftarrow \mathbf{0}$

\While{$\|\boldsymbol{\omega}^l - \boldsymbol{\omega}^l_{\text{prev}}\|_2 > \epsilon$ 
\textbf{or} 
$\|\boldsymbol{\omega}^u - \boldsymbol{\omega}^u_{\text{prev}}\|_2 > \epsilon$}{

    \textbf{Update:} $\boldsymbol{\omega}^l_{\text{prev}} \leftarrow 
    \boldsymbol{\omega}^l$,\quad 
    $\boldsymbol{\omega}^u_{\text{prev}} \leftarrow \boldsymbol{\omega}^u$

    \ForEach{$\omega_i \in \boldsymbol{\omega}$ \textnormal{(in parallel)}}{

        \textbf{Solve} in parallel:
        \begin{subequations}\label{sbt_equations}
        \begin{align}
            \omega_i^{l*} &:= \min \quad \omega_i \label{sbt:min}\\
            \omega_i^{u*} &:= \max  \quad \omega_i \label{sbt:max}
        \end{align}
        \text{subject to}\hspace{120pt}
        \begin{align}
            &f(\mathbf{x}, \mathbf{z}) \geq 
            f_{\text{nlp}}(\mathbf{x}^*, \mathbf{z}^*) 
            \label{sbt:obj_bound}\\
            &g_{\text{aff}}(\boldsymbol{\omega}, \mathbf{s}) = 0 
            \label{sbt:eq}\\
            &h_{\text{cvx}}(\boldsymbol{\omega}, \mathbf{x}, \mathbf{z}, 
            \mathbf{s}) \leq 0 
            \label{sbt:limits}\\
            &h_{\text{mc}}(\boldsymbol{\omega}, \mathbf{s}) \leq 0 
            \label{sbt:mc}\\
            &\boldsymbol{\omega}^l_{\text{prev}} \leq \boldsymbol{\omega} 
            \leq \boldsymbol{\omega}^u_{\text{prev}} 
            \label{sbt:bounds}\\
            &\mathbf{x} \in \{0,1\}^{n_x} 
            \label{sbt:binary}\\
            &\mathbf{z} \in \mathbb{Z}_{+}^{n_z} 
            \label{sbt:integer}
        \end{align}
        \end{subequations}
    }

    \textbf{Update:} $\boldsymbol{\omega}^l \leftarrow 
    \boldsymbol{\omega}^{l*}$,\quad 
    $\boldsymbol{\omega}^u \leftarrow \boldsymbol{\omega}^{u*}$
}

\KwOut{Tightened bounds $\boldsymbol{\omega}^l$, $\boldsymbol{\omega}^u$}
\end{algorithm}

As illustrated in Fig.~\ref{fig:opt_engine}, reformulating the \MIBLP{} as an \MIQCP{} enables the implementation of SBT through Algorithm~\ref{alg:sbt}.
The algorithm computes tight bounds for all bilinear and lifted variables represented by the vector $\boldsymbol{\omega}$, where $\boldsymbol{\omega}=\{s, u_1, u_2\}$ for every bilinear term.
The algorithm takes as input initial loose bounds on $\boldsymbol{\omega}$ and a bound-tightening tolerance $\epsilon$.
Before applying SBT, we impose an objective bound obtained from Algorithm~\ref{alg:penalty_heuristic}, as defined in~\eqref{sbt:obj_bound}.
After initializing all variable bounds, the algorithm solves two optimization problems for each variable $\omega_i$: one minimizing $\omega_i$ and the other maximizing $\omega_i$. These optimization problems can be solved independently and in parallel.
The resulting lower and upper bounds are iteratively refined by repeatedly solving~\eqref{sbt_equations} until the bound improvement falls below the tolerance $\epsilon$.
\eqref{sbt:eq} represents the affine AC network equations, while \eqref{sbt:limits} captures the convex operational limits and other linear constraints originally defined in~\eqref{MIBLP_c}.
\eqref{sbt:mc} enforces the McCormick envelopes associated with the bilinear terms, \eqref{sbt:bounds} defines the current bounds on the bilinear and lifted variables.

We apply the tightened bounds $(\mathbf{s}^l, \mathbf{s}^u, \mathbf{y}^l, \mathbf{y}^u)$ obtained from Algorithm~\ref{alg:sbt} to construct the tightened problem \hpmiblp{} in~\eqref{MIBLP_grid_eco}.
Further, once Algorithm~\ref{alg:penalty_heuristic} identifies an integer- and nonlinear-feasible solution, we warm-start the \hpmiblp{} in~\eqref{MIBLP_grid_eco} using the incumbent solution $(\hat{\mathbf{x}},\hat{\mathbf{z}},\hat{\mathbf{y}})$. 
\begin{subequations}\label{MIBLP_grid_eco}
    \begin{align}
        \hat{\textbf{\emph{P}}}_{\text{miblp}}:\max\quad & f(\mathbf{x}, \mathbf{z}) \label{MIBLP_grid_eco_a} \\
        \text{s.t.}\quad 
        & \eqref{MIBLP_b}, \eqref{MIBLP_c}, \eqref{MIBLP_e} \label{MIBLP_grid_eco_b} \\
        & \mathbf{y}^l \le \mathbf{y} \le \mathbf{y}^u \label{MIBLP_grid_eco_c} \\
        & \mathbf{s}^l \le \mathbf{s} \le \mathbf{s}^u \label{MIBLP_grid_eco_d} \\
        & \mathbf{x} \in \{0,1\}^{n_x} \label{MIBLP_grid_eco_e} \\
        & \mathbf{z} \in \mathbb{Z}_{\ge 0}^{n_x} \label{MIBLP_grid_eco_f}
    \end{align}
\end{subequations}

\section{Case Studies}\label{casestudy}
In this section, we evaluate the performance of Grid-ECO in four feeders in the City of Seattle: two are located in the downtown area, and the other two are in residential neighborhoods.
We solve the EVCS charger placement using three different approaches.  
Approach 1, \GurobiOff{}, solves the original \pmiblp{} in \eqref{MIBLP} using the off-the-shelf global optimizer, Gurobi. 
Approach 2, \GurobiSBT{}, solves the same \pmiblp{} using the incumbent generated by Algorithm~\ref{alg:penalty_heuristic} as a warm start and applies bound tightening only to the bilinear variables.
Approach 3, \GurobiSBTP{}, solves the tightened \hpmiblp{} in~\eqref{MIBLP_grid_eco}, also using the warm-start incumbent from Algorithm~\ref{alg:penalty_heuristic}, while applying bound tightening to both the bilinear variables and the lifted McCormick variables through Algorithm~\ref{alg:sbt}.

\subsection{Parameter Settings}
For the transportation model, we use multiple data sources~\cite{bureau2024decennial,roers20242021,bureau2012american} to characterize off-street parking availability, demographic attributes, and EV adoption rates in Seattle, WA. We set the policy weighting parameter $\alpha$ to 0.85 to balance efficiency-driven utilization and equity-oriented accessibility~\cite{lezcano2025siting}. We set the city-wide charger target $b$ to 3,085, corresponding to five charging ports per 1,000 households across the study area.

For the grid model, we assume a Level-2 charger rating of 7.2~kW~\cite{marsh2023energysage} with a power factor of 0.985~\cite{nerc2024ev}. We set the average installation cost to \$8{,}600 per charger~\cite{lyons2023current}. Because actual land acquisition costs are unavailable, we estimate them based on parking availability. The land acquisition costs used in the case studies are available in the repository~\cite{GridECO2026}. We set the total installation budget to \$2.5M and \$3M for Feeders D1 and D2, respectively, and to \$2M and \$5M for Feeders R1 and R2, respectively. While these parameters reflect a realistic urban deployment scenario, utilities and planners can readily adjust them to match local economic and operational conditions. To prevent spatial clustering, we set the service radius to 100~m in the downtown area and 200~m in the residential neighborhood.

We solve the \GurobiOff{}, \GurobiSBT{}, \GurobiSBTP{}, and all \MIQCP{} using the Gurobi solver~\cite{gurobi13manual} (version~13.0.0). We solve the \PNLP{} using IPOPT (version~3.14.17) with the HSL MA57 linear solver~\cite{HSL2007}. All optimization experiments are conducted on the Vermont Advanced Computing Core (VACC) high-performance computing (HPC) cluster~\cite{uvm_vacc}, using a 130-core processor with 200~GB of RAM.

\subsection{Study Region}
\label{sec:study_region}
\begin{figure}[htbp]
    \centering
    \includegraphics[width=1\linewidth]{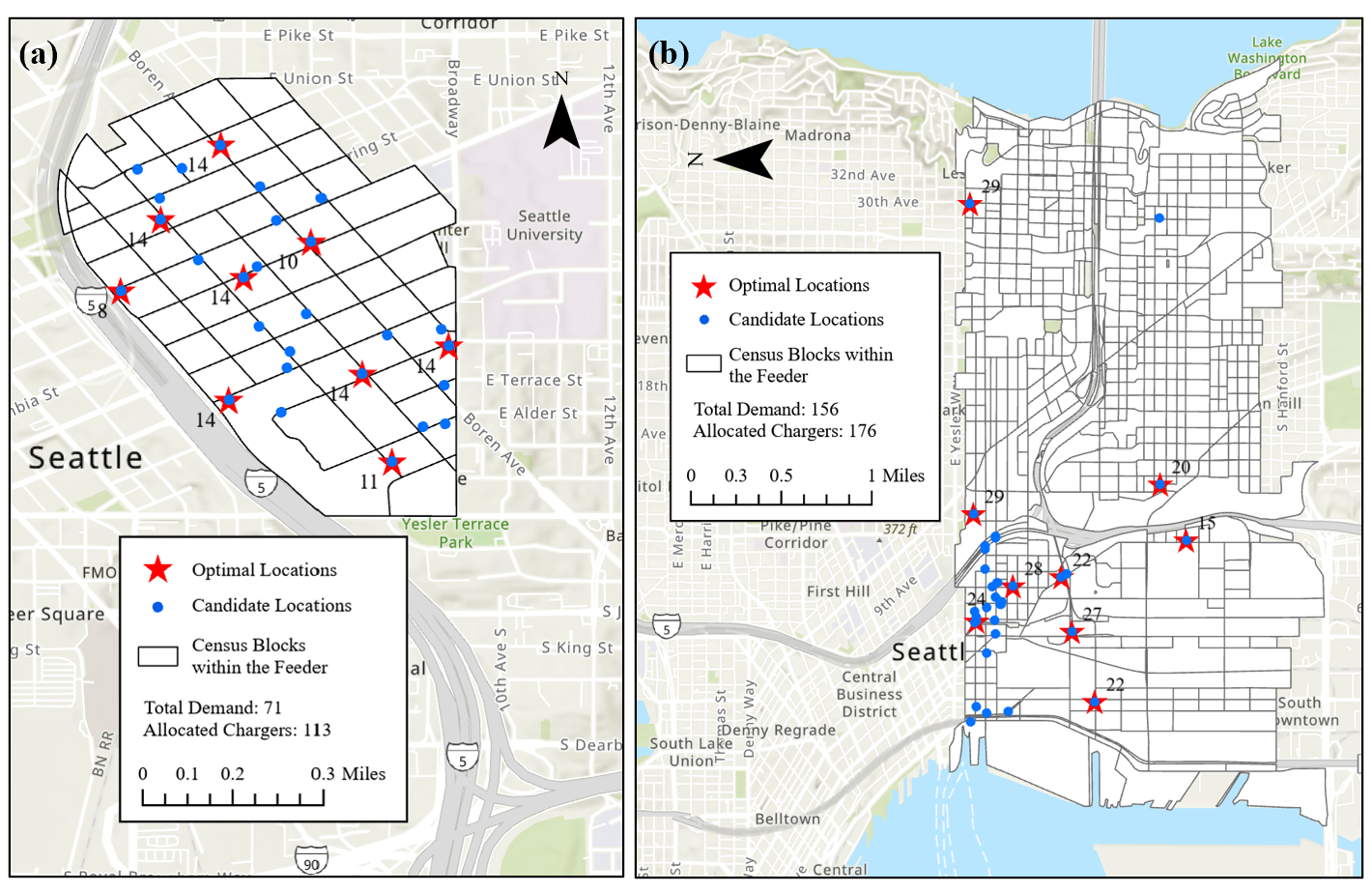}
    \caption{Distribution of candidate and optimal EVCS locations, along with installed charger counts obtained using \GurobiSBTP{} (see Table~\ref{tab:comparison}), for Feeder D1 (a) and Feeder D2 (b) in the downtown region. Blue circles denote candidate locations, red stars denote optimal locations, and numbers indicate the allocated chargers at each location.}
    \label{fig:results_downtown}
\end{figure}

\begin{figure}[!htbp]
    \centering
    \includegraphics[width=1\linewidth]{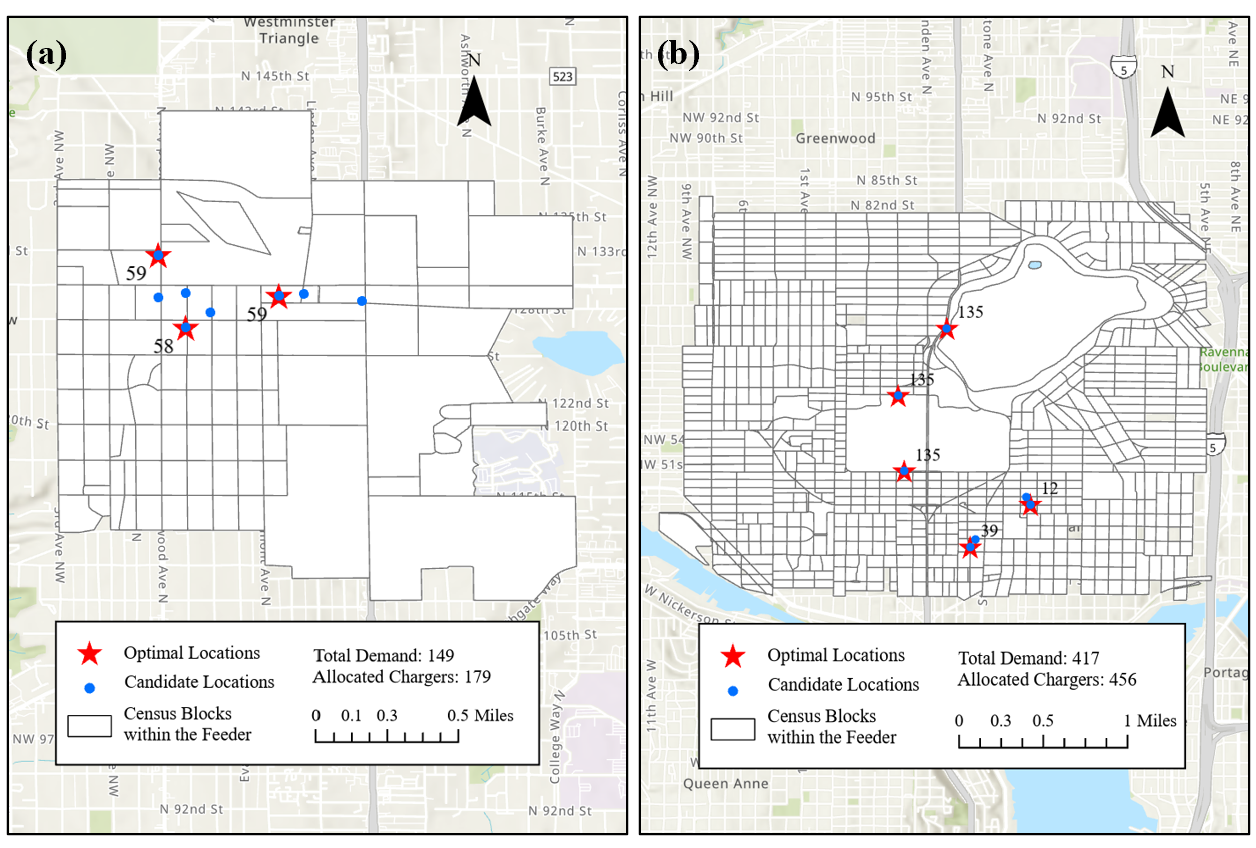}
    \caption{Distribution of candidate and optimal EVCS locations, along with installed charger counts obtained using \GurobiSBTP{} (see Table~\ref{tab:comparison}), for Feeder R1 (a) and Feeder R2 (b) in the residential neighborhood region. Blue circles denote candidate locations, red stars denote optimal locations, and numbers indicate the allocated chargers at each location.}
    \label{fig:results_rural}
\end{figure}

We evaluate Grid-ECO on two study regions in the City of Seattle: i) a downtown region (Fig.~\ref{fig:results_downtown}) and ii) a residential neighborhood region (Fig.~\ref{fig:results_rural}). For each region, we consider two distribution feeders. Due to the unavailability of real feeder data, we use representative taxonomy synthetic feeders~\cite{schneider2008modern}, as summarized in Table~\ref{tab:study_region}.

The feeders are denoted as Downtown (D1), Downtown (D2), Residential (R1), and Residential (R2), and these labels are used consistently throughout Tables~\ref{tab:presolve_pnlp}, \ref{tab:presolve_sbt}, and~\ref{tab:comparison}. Using the charging demand estimation model described in Section~\ref{sec:charging_demand_modeling}, Grid-ECO estimates the minimum charger demand for each feeder (see Table~\ref{tab:study_region}). Following the location selection and prioritization procedure described in Section~\ref{prioritize}, Grid-ECO also identifies the candidate EVCS locations for each feeder (see Table~\ref{tab:study_region}). 
\begin{table}[!htbp]
\centering
\caption{Study feeders and corresponding properties.}
\label{tab:study_region}
\footnotesize
\setlength{\tabcolsep}{4pt}
\begin{tabular}{llrrr}
\toprule
Case & Feeder & Buses & Demand.$^{a}$ & Candidate.$^{b}$ \\
\midrule
Downtown (D1)    & \texttt{R3-12.47-2} & 863  & 71  & 27 \\
Downtown (D2)    & \texttt{R5-35.00-1} & 2008 & 156 & 32 \\
Residential (R1) & \texttt{R1-25.00-1} & 830  & 149 & 8   \\
Residential (R2) & \texttt{R5-12.47-2} & 1980 & 417 & 7   \\
\bottomrule
\end{tabular}

\vspace{2pt}
{\footnotesize\raggedright
$^{a}$ Estimated minimum charger demand.\quad
$^{b}$ Number of candidate EVCS locations.\par}
\end{table}

\subsection{Comparison of Solution Approaches and Analysis}
\begin{table*}[!t]
\centering
\caption{Performance comparison of solution approaches for solving \pmiblp{}}
\label{tab:comparison}
\renewcommand{\arraystretch}{1.02}
\setlength{\tabcolsep}{7pt}
\begin{threeparttable}
\resizebox{\textwidth}{!}{%
\begin{tabular}{@{} l r r l r r r r r r @{}}
\toprule
\textbf{Case}
    & \textbf{Variables}\tnote{a}
    & \textbf{Constraints}
    & \textbf{Approach}
    & \textbf{Objective}
    & \textbf{Time (s)}\tnote{c}
    & \textbf{\# sBnB Nodes}
    & \textbf{Gap (\%)}
    & \textbf{EVCS Sites}
    & \textbf{Chargers} \\
\midrule
\multirow{3}{*}{Downtown (D1)}
    & \multirow{3}{2.2cm}{\centering 5{,}526 \\ {\scriptsize (5472c\,/\,27i\,/\,27b)}}
    & 7{,}865 & \GurobiOff{}  & $-$  & \textcolor{red}{43{,}200}\tnote{\textdagger} & $-$ & $-$ & $-$ & $-$ \\
  & & 7{,}865      & \GurobiSBT{}  & 4.15 & 97.81 & 182 & 0.00 & 9 & 113 \\
  & & 8{,}081      & \GurobiSBTP{} & 4.15 & 57.54 & 3   & 0.00 & 9 & 113 \\
\midrule
\multirow{3}{*}{Downtown (D2)}
    & \multirow{3}{2.2cm}{\centering 9{,}924 \\ {\scriptsize (9860c\,/\,32i\,/\,32b)}}
    & 14{,}054 & \GurobiOff{}  & 6.75\tnote{b} & \textcolor{red}{43{,}200}\tnote{\textdagger} &  555{,}531 & 1.39 & $-$ & $-$ \\
  & & 14{,}054 & \GurobiSBT{}  & 6.75 & 888.11 & 2{,}649 & 0.00 & 9 & 216 \\
  & & 14{,}310 & \GurobiSBTP{} & 6.75 & 279.11 & 127 & 0.00 & 9 & 216 \\
\midrule
\multirow{3}{*}{\shortstack[l]{Residential (R1)}}
    & \multirow{3}{2.2cm}{\centering 5{,}122 \\ {\scriptsize (5106c\,/\,8i\,/\,8b)}}
    & 6{,}797 & \GurobiOff{}  & $-$   & \textcolor{red}{43{,}200}\tnote{\textdagger} & $-$ & $-$ & $-$ & $-$ \\
  & & 6{,}797     & \GurobiSBT{}  & 21.35 & 5.96 & 1 & 0.00 & 3 & 176 \\
  & & 6{,}861     & \GurobiSBTP{} & 21.35 & 4.31 & 1 & 0.00 & 3 & 176 \\
\midrule
\multirow{3}{*}{\shortstack[l]{Residential (R2)}}
    & \multirow{3}{2.2cm}{\centering 9{,}964 \\ {\scriptsize (9950c\,/\,7i\,/\,7b)}}
    & 13{,}784 & \GurobiOff{}  & $-$ & \textcolor{red}{43{,}200}\tnote{\textdagger} & $-$ & $-$ & $-$ & $-$ \\
  & & 13{,}784 & \GurobiSBT{}  & 64.99 & 670.51 & 1{,}534 & 0.00 & 5 & 456 \\
  & & 13{,}840 & \GurobiSBTP{} & 64.99 & 753.22 & 1{,}358 & 0.00 & 5 & 456 \\
\bottomrule
\end{tabular}%
}
\begin{tablenotes}[flushleft]
    \footnotesize
    \item[a] Variable breakdown: \textbf{c}\,=\,continuous, \textbf{i}\,=\,integer,
             \textbf{b}\,=\,binary.
    \item[b] Incumbent objective $=6.75$; best bound $=6.85$ (gap reported as Gurobi default).
    \item[c] Solver time only; excludes \PNLP{} (Table~\ref{tab:presolve_pnlp}) and SBT
             presolve (Table~\ref{tab:presolve_sbt}).
    \item[\textdagger] Time limit of 43{,}200\,s (12\,hours) reached.
\end{tablenotes}
\end{threeparttable}
\end{table*}

Table~\ref{tab:presolve_pnlp} summarizes the performance of the \PNLP{} heuristic, including objective value, solution time, number of penalty iterations (epochs), selected EVCS locations, and installed chargers. Compared with the optimal Grid-ECO (\GurobiSBTP{}) solutions reported in Table~\ref{tab:comparison}, the \PNLP{} solutions are suboptimal but provide high-quality feasible incumbents that effectively warm-start the solution of \hpmiblp{}.
\begin{table}[!htbp]
\centering
\caption{Presolve: \PNLP{} heuristic.}
\label{tab:presolve_pnlp}
\renewcommand{\arraystretch}{1.02}
\setlength{\tabcolsep}{5pt}
\begin{threeparttable}
\begin{tabular}{@{} l r r r r r @{}}
\toprule
\textbf{Case} & \textbf{Objective} & \textbf{Time (s)} & \textbf{Epoch}\tnote{a} & \textbf{EVCS Sites} & \textbf{Chargers} \\
\midrule
Downtown (D1)           & 2.66  & 60.70 & 19 & 7  & 71  \\
Downtown (D2)            & 4.87    & 21.36 & 5 & 10 & 156 \\
Residential (R1)  & 18.51 & 9.91  & 17 & 5  & 149 \\
Residential (R2)  & 59.43   & 11.53    & 8 & 4 & 417 \\
\bottomrule
\end{tabular}
\begin{tablenotes}[flushleft]
\footnotesize
\item[a] Epoch\,=\,penalty iterations to an integer-feasible solution.
\end{tablenotes}
\end{threeparttable}
\end{table}
Table~\ref{tab:presolve_sbt} compares two bound-tightening strategies: i) \texttt{SBT}, which tightens only bilinear variables, and ii) \texttt{SBT$^{+}$}, which tightens both bilinear and lifted McCormick variables. Across the four cases, \texttt{SBT} tightens between 590 and 1,540 variables and requires 83.71--802.07\,s, whereas \texttt{SBT$^{+}$} tightens 1,234--3,528 variables and requires 138.09--1,667.01\,s. While \texttt{SBT$^{+}$} requires additional presolve effort, it yields tighter variable bounds and stronger relaxations, leading to improved solver performance in Table~\ref{tab:comparison}. Since the bound-tightening procedures are highly parallelizable and were executed on 130 CPU cores, their runtime can be further reduced on systems with additional computational resources.
\begin{table}[!htbp]
\centering
\caption{Presolve: Bound Tightening using SBT.}
\label{tab:presolve_sbt}
\renewcommand{\arraystretch}{1.01}
\setlength{\tabcolsep}{19pt}
\begin{threeparttable}
\begin{tabular}{@{} l l r r @{}}
\toprule
\textbf{Case} & \textbf{Strategy}\tnote{a} & \textbf{Variables}\tnote{b} & \textbf{Time (s)}\tnote{c} \\
\midrule
\multirow{2}{*}{Downtown (D1)}
   & \texttt{SBT}      & 590     & 83.71  \\
 & \texttt{SBT$^{+}$} & 1{,}234 & 138.09 \\
\midrule
\multirow{2}{*}{Downtown (D2)}
   & \texttt{SBT}       & 1540 & 802.07 \\
 & \texttt{SBT$^{+}$}  & 3528 & 1667.01 \\
\midrule
\multirow{2}{*}{Residential (R1)}
   & \texttt{SBT}      & 618  & 94.32 \\
 & \texttt{SBT$^{+}$}  & 1302 & 274.35 \\
\midrule
\multirow{2}{*}{Residential (R2)}
   & \texttt{SBT}      & 1254 & 619.60 \\
 & \texttt{SBT$^{+}$}  & 2798 & 1291.15 \\
\bottomrule
\end{tabular}
\begin{tablenotes}[flushleft]
\footnotesize
\item[a] \texttt{SBT}: Strategy~A (bilinear variables only); \texttt{SBT$^{+}$}: Strategy~B (bilinear and lifted variables).
\item[b] Number of variables tightened.
\item[c] Bound-tightening time for all variables.
\end{tablenotes}
\end{threeparttable}
\end{table}

Table~\ref{tab:comparison} compares the four solution approaches for \pmiblp{} and \hpmiblp{} instances ranging from 5,122 variables (Residential~R1) to 9,964 variables (Residential~R2). The primary finding is the importance of presolve. \GurobiOff{} fails to certify optimality within the 12-hour time limit for all four cases and finds no feasible solution in three of them, whereas both \GurobiSBT{} and \GurobiSBTP{} achieve a reported optimality gap of $0.00\%$ in every case.

Relative to \GurobiSBT{}, \GurobiSBTP{} reduces the number of explored \sbnb{} nodes from 182 to 3 on D1 and from 2{,}649 to 127 on D2---up to a $98\%$ reduction---while decreasing solver time by 41\% and 69\%, respectively. These improvements arise from tighter relaxations obtained by additionally tightening lifted McCormick variables. The improved search performance comes at the cost of additional presolve effort (Table~\ref{tab:presolve_sbt}); therefore, the overall wall-clock benefit depends on the relative cost of presolve and branch-and-bound. The benefit is most pronounced on D1 and D2, where the reduction in explored nodes is substantial. In contrast, for R2 the node reduction is modest (1{,}534 to 1{,}358, approximately 11\%). These results suggest that SBT$^{+}$ may be particularly beneficial for larger instances where branch-and-bound dominates the overall solution effort.

\FloatBarrier


\section{Conclusion}
In this paper, we introduced Grid-ECO, an optimization framework for allocating and deploying EV chargers while satisfying nonlinear three-phase unbalanced AC network constraints.
We evaluated the framework on four large-scale distribution feeder models representing downtown and residential regions of Seattle, with problem sizes ranging into the thousands, with both continuous and integer variables.
The main findings are as follows:

\begin{itemize}

\item \textbf{Exact reformulation of unbalanced AC grid physics.} Grid-ECO exactly formulates the EVCS placement problem, with unbalanced AC grid physics, as a nonconvex \MIBLP{}, enabling the use of a spatial branch and bound global algorithm.


\item \textbf{Near-zero optimality gap for exact problem.} 
Grid-ECO develops presolve routines (\GurobiSBTP{}) that solve four instances of the EVCS placement problem to a $0.00\%$ optimality gap, with a maximum solver time of 753s.
Off-the-shelf Gurobi (\GurobiOff{}) failed to certify optimality within the 12-hour time limit and did not find a feasible solution in three of the four test cases.

\item \textbf{Lifted-variable bound tightening.} Applying SBT to both bilinear and lifted  variables (\GurobiSBTP{}) produced tighter relaxations and substantially reduced \sbnb{} search effort. Compared with standard SBT (\GurobiSBT{}), \GurobiSBTP{} reduced the number of explored nodes by up to 98\% and solver time by up to 69\%. The improvement was most significant for instances where branch-and-bound dominated the overall solution effort, demonstrating the effectiveness of lifted-variable bound tightening for challenging nonconvex \MIBLP{}s.

\end{itemize}

\section{Acknowledgement} 
\vspace{-0.5em}
\noindent
This research was supported by the U.S. Department of Energy Vehicle Technologies Office (Grant No. DEEE0010640) and the Alfred P. Sloan Foundation (Grant No. 2024-22563).

\bibliographystyle{IEEEtran}
\bibliography{reference}

\end{document}